\documentclass[12pt]{iopart}

\usepackage{iopams}  
\usepackage[T1]{fontenc}
\usepackage[latin9]{inputenc}
\usepackage{verbatim}
\usepackage{units}
\usepackage{amstext}
\usepackage{graphicx}
\usepackage{subfigure}
\usepackage{esint}
\begin{document}

\title{Kinetic modelling of runaway electron avalanches in tokamak plasmas}

\author{E. Nilsson$^{1}$, J. Decker$^{2}$, Y. Peysson$^{1}$,
R.S Granetz$^{3}$, \\F. Saint-Laurent$^{1}$, M. Vlainic$^{4,5}$}

\address{$^{1}$CEA, IRFM, F-13108, Saint-Paul-lez-Durance, France}
\address{$^{2}$Ecole Polytechnique F\' ed\'erale de Lausanne (EPFL), Centre de Recherches en Physique des Plasmas (CRPP), CH-1015 Lausanne, Switzerland}
\address{$^{3}$MIT Plasma Science and Fusion Center, Cambridge,
Massachusetts 02139, USA}
\address{$^{4}$Department of Applied Physics, Ghent University,
B-9000, Ghent, Belgium}
\address{$^{5}$Institute of Plasma Physics AS CR, 18200 Prague,
Czech Republic}

\ead{emelie.nilsson@cea.fr}
\vspace{10pt}
\begin{indented}
\item[]13 March 2015
\end{indented}

\begin{abstract}
Runaway electrons can be generated in tokamak plasmas if the accelerating
force from the toroidal electric field exceeds the collisional drag
force owing to Coulomb collisions with the background plasma. In ITER,
disruptions are expected to generate runaway electrons mainly through
knock-on collisions \cite{hen07}, where enough momentum can be transferred
from existing runaways to slow electrons to transport the latter beyond
a critical momentum, setting off an avalanche of runaway electrons.
Since knock-on runaways are usually scattered off with a significant
perpendicular component of the momentum with respect to the local
magnetic field direction, these particles are highly magnetized. Consequently,
the momentum dynamics require a full 3-D kinetic description, since
these electrons are highly sensitive to the magnetic non-uniformity
of a toroidal configuration. For this purpose, a bounce-averaged knock-on
source term is derived. The generation of runaway electrons from the
combined effect of Dreicer mechanism and knock-on collision process
is studied with the code LUKE, a solver of the 3-D linearized bounce-averaged
relativistic electron Fokker-Planck equation \cite{dec04a}, through
the calculation of the response of the electron distribution function
to a constant parallel electric field. The model, which has been successfully
benchmarked against the standard Dreicer runaway theory now describes
the runaway generation by knock-on collisions as proposed by Rosenbluth
\cite{ros97}. This paper shows that the avalanche effect can be important
even in non-disruptive scenarios. Runaway formation through knock-on
collisions is found to be strongly reduced when taking place off the
magnetic axis, since trapped electrons can not contribute to the runaway
electron population. Finally, the relative importance of the avalanche
mechanism is investigated as a function of the key parameters for
runaway electron formation, namely the plasma temperature and the
electric field strength. In agreement with theoretical predictions,
the LUKE simulations show that in low temperature and electric field
the knock-on collisions becomes the dominant source of runaway electrons
and can play a significant role for runaway electron generation, including
in non-disruptive tokamak scenarios.
\end{abstract}

%
%
%
%
%
\section{Introduction}

Runaway electrons have been observed in magnetic confinement fusion
experiments during the operation of tokamaks \cite{wes89}. They are
also encountered in nature in solar flares and electric discharges
associated with thunderstorms \cite{tav11}. The dynamics of electrons
in a plasma is governed by the balance between acceleration in an
electric field and collisions with the plasma particles. Collisional
friction forces acting on the electrons reach a global maximum at
the thermal velocity ($v_{th}$) and decrease for higher velocities.
In the presence of a strong toroidal electric field ($E$) collisional
drag may consequently be too weak to counteract the acceleration of
electrons, which may result in continuously accelerated electrons,
known as runaway electrons. If no other loss mechanisms than the collisional
drag are present \cite{con75}, runaway electrons may be generated
if the electric field exceeds the critical field \cite{dre59}

\begin{equation}
E_{c}=\frac{n_{e}e^{3}\ln\Lambda}{4\pi\varepsilon_{0}^{2}m_{0}c^{2}},\label{eq:ec}
\end{equation}
where $n_{e}$ is the electron density, $m_{0}$ is the electron rest
mass, $c$ is the speed of light, $e$ is the elementary charge, and
$\ln\Lambda$ is the Coulomb logarithm. The acceleration by a DC field
of electrons that diffuse via small angle collisions beyond the critical
momentum ($p_{c}$), defined as the minimum momentum for which collisions
are too weak to prevent acceleration of the electrons by the electric
field to even higher energies, is referred to as the Dreicer mechanism
\cite{dre59}. In addition, these relativistic electrons can undergo
close collisions with bulk electrons and transfer part of their momentum
so that also the target electrons may get kicked into the runaway
momentum region, while the momentum of the primary electrons remains
above the critical momentum. These knock-on collisions can therefore
lead to multiplication of the number of runaway electrons, commonly
referred to as runaway avalanche \cite{ros97}. 

Various methods to mitigate the formation of runaway electrons in
tokamak plasmas are based on either increasing the plasma density
and thereby $E_{c}$ by so-called massive gas injection (MGI) \cite{leh11},
or on deconfining the runaway electrons before they can reach too
high energy, by the means of resonant magnetic perturbations (RMP)
\cite{leh09}. Even though such mitigation methods have been demonstrated
in present tokamak experiments, they might not provide a solution
for large tokamaks like ITER \cite{hol15}. Therefore the formation
of the runaway electron population is a topic in urgent need of investigation.

Intense beams of highly energetic runaway electrons can form in tokamaks
during plasma disruptions, fast unstable events that lead to a sudden
loss of plasma confinement. If runaway electrons strike the first
wall of the tokamak vacuum vessel the local energy deposition can
cause significant damage \cite{hen07}. Regardless of the mechanisms
that lead to the onset of a major disruption, the post-disruption
phases usually have similar time evolution \cite{hol11}. They start
with a fast cooling of the plasma typically associated with either
intense radiative losses or ergodisation of the magnetic flux surfaces
\cite{bon91}, referred to as the thermal quench, which occurs on
a time scale on the order of a millisecond. Consequently the plasma
resistivity $\rho$, which scales with the temperature as $T^{-3/2}$,
increases rapidly. The toroidal electric field is proportional to
the resistivity and increases dramatically in order to maintain the
local current density. The resistive current diffusion is enhanced
by the reduction of the plasma conductivity, such that the plasma
current decays progressively. Yet, the current decay occurs over a
much longer time scale. In this process, a fraction of the pre-disruptive
plasma current is carried by runaway electrons.

Disruptions are interesting but complex processes for studying the
birth of runaway electrons, since they include magnetohydrodynamic
(MHD) instabilities, anomalous transport and complex evolution of
the magnetic field topology \cite{ned08}. However, the generation
of runaway electrons does not necessarily require the extreme conditions
found in disruptions. In low density plasmas, the electric field can
exceed the critical electric field also during the current flattop
in a quiescent plasma, free of equilibrium transients, or during current
ramp up or ramp down. An advantage of studying runaway formation in
these so called non-disruptive scenarios is that the key parameters
for the runaway electron mechanisms, mainly the electric field strength,
electron density and temperature, can be better diagnosed than during
disruptions. Runaway electrons have been detected in non-disruptive
scenarios in several of the existing tokamaks \cite{gra14}. Quiescent
plasmas with nested magnetic flux surfaces are therefore more suitable
for studying the formation of runaway electrons. In this work the
formation of runaway electrons generated from the combined effect
of Dreicer and knock-on collision mechanisms is studied with the code
LUKE, a solver of the 3-D (one spatial and two momentum dimensions)
linearized bounce-averaged relativistic electron Fokker-Planck equation
\cite{dec04a}. The code LUKE handles arbitrary shapes of the flux
surfaces, but in this work the magnetic flux surfaces are assumed
to remain circular and concentric as in the Tore Supra tokamak. They
are assumed to remain intact throughout the runaway formation process,
an assumption that would be too restrictive for the thermal quench
in disruptive scenarios. 

Modelling the evolution of the temperature and electric field in disruptions
would require a proper description of the thermal quench including
radiative or convective loss mechanisms and MHD instabilities. The
coupling of a kinetic code capable of handling \mbox{3-D} magnetic topologies
and open field lines with a fluid code such as JOREK \cite{huy07}
would be necessary for such a purpose, but is beyond the scope of
this work. The kinetic modelling of the formation of runaway electrons
is therefore done for non-disruptive scenarios as found in the current
flattop with constant electric field and plasma temperature. With
the restrictions of disruption modelling in mind, the objective of
this work is to study the formation of runaway electrons in non-disruptive
scenarios owing to the combined effect of Dreicer and knock-on collisions
with a fast solver for the electron distribution function, in order
to make predictions for the birth of runaway electrons in tokamak
experiments. 

The LUKE code has previously been used for current drive and Dreicer
runaway calculations. The model uses a relativistic collision operator
for small angle collisions and a recently added description of the
large angle (knock-on) collisions leading to the avalanche effect,
which enables a description of the full 2-D momentum dynamics of the
runaway population. Runaway electrons generated via knock-on collisions
are typically scattered off with a significant perpendicular component
of the momentum with respect to the local magnetic field direction.
In a non-uniform magnetic field configuration, highly magnetized electrons
could be subject to magnetic trapping effects resulting in reduced
runaway electron growth rate off the magnetic axis in comparison to
estimates obtained for a cylindric geometry. Such toroidicity effects
are studied by implementing a 2-D kinetic description of the knock-on
momentum dynamics, including the momentum dynamics both perpendicular
and parallel to the magnetic field lines. 

Knock-on collisions are included in the kinetic equation through a
source term from Ref.~\cite{ros97}, implemented along with a sink
term to ensure a particle conserving form of the process. The bounce-averaged
knock-on source term is presented in Sec.~\ref{sec:Knock-on-collisions}.
In Sec.~\ref{sec:Effect-of-toroidicity} the effect of magnetic field
non-uniformity is investigated. The role played by the magnetic mirror
force on the runaway population off the magnetic axis, owing to a
reduction in Dreicer growth rate as well as the high magnetization
of the knock-on electrons, is described. Finally, in Sec.~\ref{sec:Importance-of-avalanche},
the relative importance of the avalanche effect compared to the Dreicer
mechanism is quantified as a function of plasma temperature and toroidal
electric field strength. The parametric dependencies of the relative
importance of the avalanche effect obtained from the numerical modelling
is related to both analytic predictions and experimental data from
runaway observations in non-disruptive scenarios from several tokamaks.
The comparison includes a low density flattop pulse from the Tore
Supra tokamak, during which suprathermal electrons are observed. The
analysis of this scenario supports recently published results \cite{gra14},
showing that runaway electron formation requires lower density than
expected from collisional theory, which suggests the existence of
additional runaway electron loss mechanisms.

\section{Knock-on collisions model\label{sec:Knock-on-collisions}}

A knock-on collision between an existing runaway electron and a slower
electron is considered. This paper follows the model from Ref.~\cite{ros97}
in which the target electron is assumed to be at rest whereas the
initial runaway travels at the speed of light in the direction of
the magnetic field. This approximation will be justified later in
this section. The target electron gains a momentum $p$ from the close
collision. As both energy and momentum must be conserved in the collision
process, the secondary electron is scattered with a pitch-angle with
respect to the direction of the incoming electron, which cosine $\xi^{*}$
is given by the relation

\begin{equation}
\xi^{*}=\sqrt{\frac{\gamma-1}{\gamma+1}},\label{eq:xi}
\end{equation}
where $\gamma=\sqrt{1+p^{2}}$ is the relativistic factor and the
momentum $p$ is normalized to $m_{0}c$.

The relativistic electron electron differential cross section derived
by Møller \cite{mol32} yields

\begin{equation}
\frac{d\sigma}{d\Omega}=r_{e}^{2}\frac{1}{p\gamma(\gamma-1)^{2}}\delta(\xi-\xi^{*}(p)),\label{eq:cs}
\end{equation}
where $r_{e}=e^{2}/(4\pi\varepsilon_{0}m_{e}c^{2})$ is the classical
electron radius. As $d\sigma/d\Omega$ decrease rapidly with momentum,
a large fraction of secondary electrons have a moderate kinetic energy
with $\gamma-1\ll1$ and are thus scattered with a large pitch-angle
characterized by $\xi^{*}\ll1$. Hence it is necessary to properly
account for the 2-D guiding-center momentum dynamics in non-uniform
magnetic field geometries, where the electrons are influenced by the
magnetic trapping effect.

The source term originally formulated in Ref.~\cite{ros97} is proportional
to both the target population, i.e. the bulk electron density $n_{e}$
and the existing runaway electron population $n_{r}$ 

\begin{eqnarray}
\mathbf{S}(\psi,p,\xi) & = & n_{e}(\psi)n_{r}(\psi)c\frac{d\sigma}{d\Omega}(p,\xi)\nonumber \\
 & = & \frac{n_{r}}{4\pi\tau\ln\Lambda}\frac{1}{p^{2}}\frac{d}{dp}\left[\frac{1}{1-\sqrt{1+p^{2}}}\right]\delta\left(\xi-\xi^{*}(p)\right),\label{eq:operator}
\end{eqnarray}
where $\psi$ is the poloidal magnetic flux surface coordinate. In
the expression above, the collision time for relativistic electrons
\begin{equation}
\tau=\frac{4\pi\varepsilon_{0}^{2}m_{e}^{2}c^{3}}{n_{e}e^{4}\ln\Lambda},\label{eq:taucc}
\end{equation}
has been introduced.

An analytic estimate of the avalanche growth rate is obtained from
integration of the knock-on source term in Eq.~\ref{eq:operator}
over the runaway region $p>p_{c}$ in momentum space, as done in the
work by Rosenbluth, which yields the following expression for $E>E_{c}$
\cite{ros97}

\begin{equation}
\frac{1}{n_{r}}\frac{\partial n_{r}}{\partial t}=\frac{1}{2\tau\ln\Lambda}\left(\frac{E}{E_{c}}-1\right).\label{eq:grRP-1}
\end{equation}

\subsection{Implementation of knock-on collisions in the LUKE code}

\begin{figure}
\centering
\includegraphics[scale=0.35]{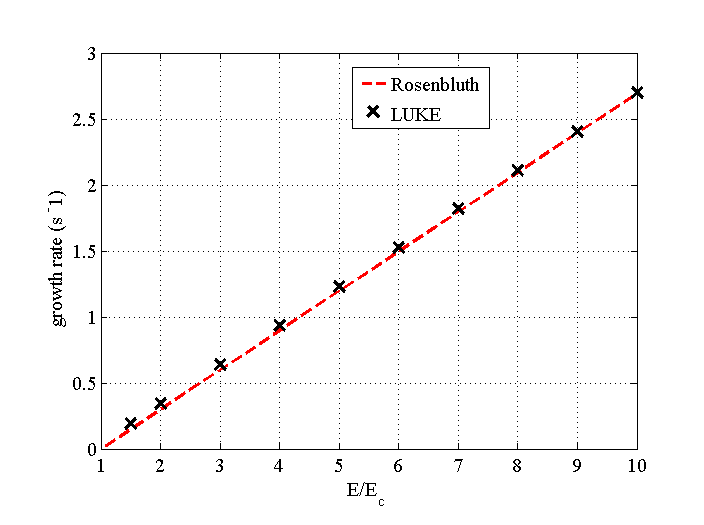}
\caption{The knock-on process in LUKE (crosses) benchmarked against analytic
growth rate in Ref.~\cite{ros97} (dashed line), when using the same
momentum space thresholds.\label{fig:Knock-on-process}}
\end{figure}

The Rosenbluth model (Eq.~\ref{eq:operator}) for the runaway generation
through knock-on collisions is implemented in the code LUKE and benchmarked
against the growth rate in Eq.~\ref{eq:grRP-1} in the case of cylindrical
geometry in Fig.~\ref{fig:Knock-on-process}, by using the same momentum
thresholds as in Ref \cite{ros97}. Electrons with a momentum larger
than $p_{re}\equiv\max[p_{c};p(E_{k}=1\,\textrm{MeV})]$ are accounted
for in the population $n_{r}$ of primary runaways for the knock-on
collision process. The numerical momentum grid boundary $p_{max}$
must be chosen to be larger or equal to $p_{re}$ and electrons leaving
the domain through the boundary remain accounted for in $n_{r}$.

To be valid, the Rosenbluth approximation requires that: (a) primary
runaways in the knock-on collision process have a velocity near the
speed of light, and (b) primary electrons have a momentum much larger
than target electrons. The condition (a) is ensured by the $1\,\textrm{MeV}$
minimum condition in $p_{re}$, which corresponds to $v/c\geq0.94$,
whereas the condition (b) is guaranteed by restricting the model to
plasmas with $T_{e}\ll1$ MeV. The Rosenbluth approximation is further
justified by the weak dependence of the knock-on source term upon
the incident electron energy in the energy range $1-100\,\textrm{MeV}$
\cite{chi98}. 

\begin{figure}
\centering
\includegraphics[scale=0.38]{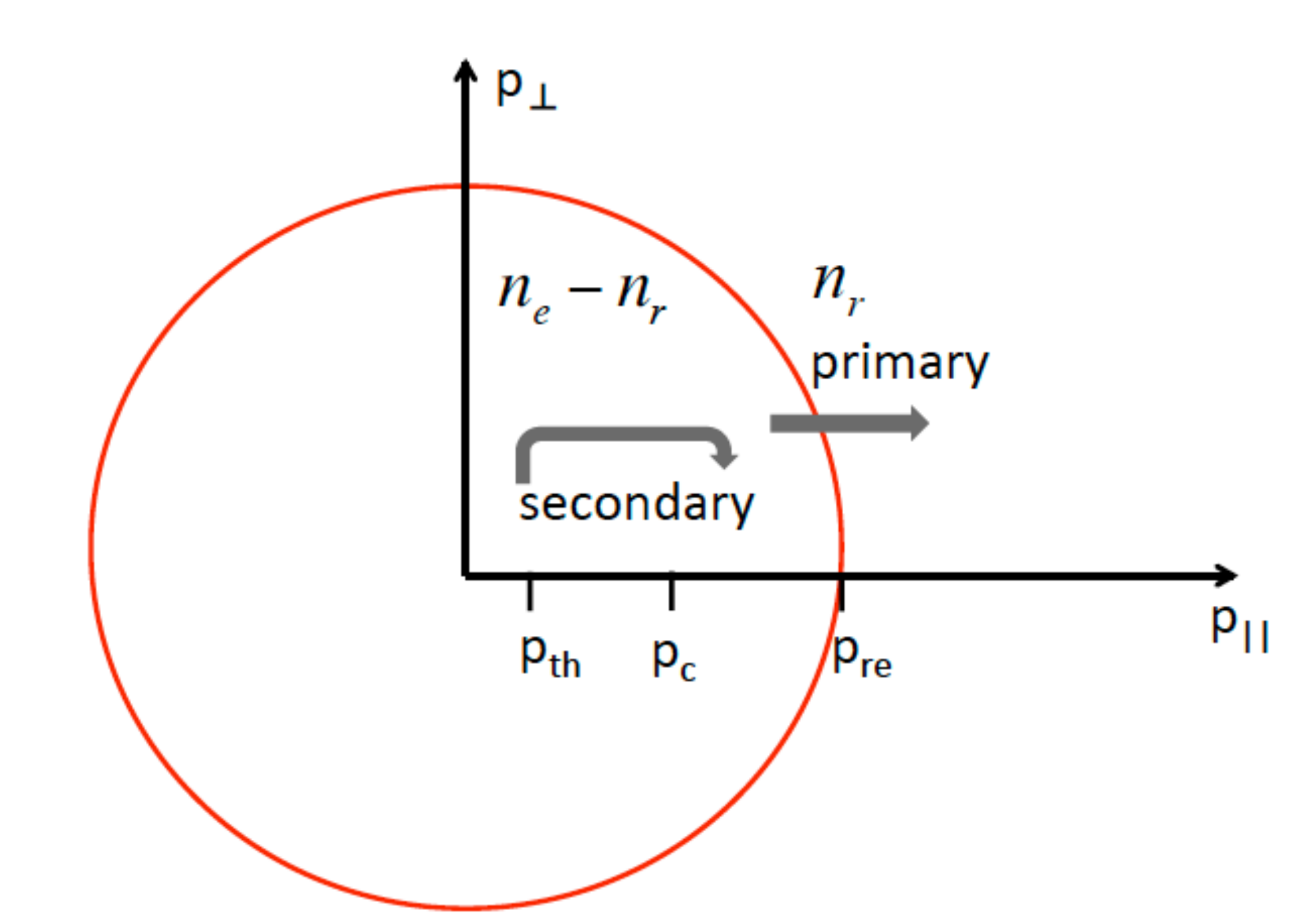}
\caption{The LUKE momentum space is divided into two separate populations:
the bulk electrons with momentum $p<p_{re}$ and the runaway electrons
$p>p_{re}$. The knock-on collisions between the populations $n_{r}$
and $n_{e}$ can lead to secondary runaway electrons. Electrons that
escape the domain $p<p_{re}$ by diffusion through $p_{re}$ contribute to the runaway electron population $n_{r}$. \label{fig:lukedomain}}
\end{figure}

The bulk electron density is defined as the integral of the bulk electron
distribution in momentum space:

\begin{equation}
\int_{_{_{0}}}^{^{p_{re}}}f(r,p)d^{3}p=n_{e}(r).
\end{equation}
The bulk and the runaway region, corresponding to $p<p_{re}$ and
$p>p_{re}$ respectively, are shown in Fig.~\ref{fig:lukedomain}.
The runaway electron population is the integral over both Dreicer
and knock-on runaway fluxes 
\[
n_{r}(t)=\int_{_{0}}^{^{t}}\left(\gamma_{D}+\int_{_{_{p_{re}}}}^{^{p_{max}}}S\, d^{3}p\right)\ dt,
\]
where $\gamma_{D}=\iint \mathbf{S_p} (\psi,p,\xi) \cdot d \mathbf{S}$
is the integral of the particle flux through the surface $p=p_{re}$.
In order to ensure conservation of number of particles in LUKE, a
sink term is implemented to compensate for the knock-on source term

\begin{equation}
\textbf{S}=\textbf{S}_{+}-<\textbf{S}_{+}>\frac{f_{M}}{<f_{M}>},
\end{equation}
where $f_{M}$ is the bulk distribution, assumed to be Maxwellian
and $<...>=\int_{_{0}}^{p_{max}}...\, d^{3}p$. The source and sink
terms ensure that the number of electrons $n_{e}+n_{r}=n_{tot}$ is
conserved. 

\subsection{Runaway electron growth rate\label{sec:Runaway-electron-growth}}

\begin{figure}
\centering
\includegraphics[scale=0.5]{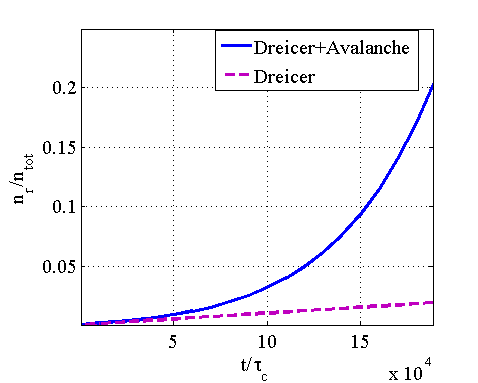}
\caption{The fraction of runaway electrons ($E/E_{c}=30$ and $T_{e}=0.5$
keV) as a function of time normalized to thermal collision time, with
and without the avalanche effect due to the knock-on collisions. \label{fig:The-runaway-population}. }
\end{figure}

The runaway electron dynamics implemented in LUKE captures the combined
effect of Dreicer and knock-on processes. The evolution of the runaway
electron population under the influence of a constant electric field
is calculated. Figure \ref{fig:The-runaway-population} shows the
evolution of a runaway electron fraction with and without knock-on
collisions. At first, there are very few runaway electrons, the knock-on
contribution becomes significant only when a primary runaway electron
population has been built up by the Dreicer effect. Then, an exponential
growth of the runaway electron population is observed - describing
the avalanche effect - and quickly becomes dominant over the Dreicer
generation. 
Both the Dreicer and avalanche mechanisms are proportional to the
bulk density $n_{e}=n_{tot}-n_{r}$, such that the runaway production
rate can be expressed in the generic form

\begin{equation}
\frac{\partial n_{r}}{\partial t}=n_{e}\left(\gamma_{D}+\gamma_{A}\right)\,\rightarrow\;\frac{1}{(n_{tot}-n_{r})}\frac{\partial n_{r}}{\partial t}=\gamma_{D}+\gamma_{A}.\label{eq:grDA}
\end{equation}
To quantify the avalanche growth rate, the avalanche term may be expressed
as $\gamma_{A}=n_{r}\bar{\gamma}_{A}$, where $\bar{\gamma}_{A}$
is an avalanche multiplication factor. Thus, Eq.~\ref{eq:grDA} becomes:

\begin{equation}
\frac{1}{(n_{tot}-n_{r})}\frac{\partial n_{r}}{\partial t}=\gamma_{D}+n_{r}\bar{\gamma}_{A}.\label{eq:grDA-1}
\end{equation}

\begin{figure}
\centering
\includegraphics[scale=0.35]{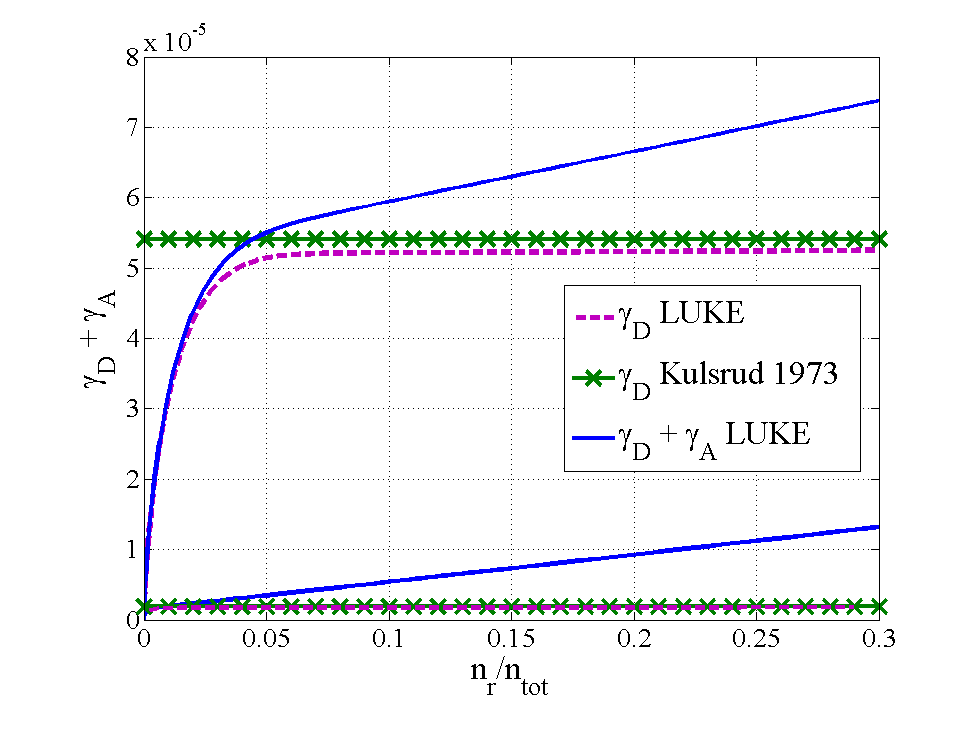}
\caption{The growth rate in constant electric field and $T_{e}=\unit{0}.5$
keV for $E/E_{c}=40$ (the curves with lower growth rate) and $E/E_{c}=60$
(curves with higher growth rate) as a function of the runaway electron
density, with and without the avalanche effect. The Dreicer contribution
is in good agreement with Kulsrud's theory \cite{kul73}. The growth
rates are normalized to the thermal collision frequency ($\nu_{th}=1/\tau(v_{th})$)\label{fig:The-growth-rate-1}}
\end{figure}

Equation \ref{eq:grDA-1} is an affine function of $n_{r}(t)$, where
the constant term is the Dreicer growth rate and the avalanche multiplication
factor is given by the slope. In Fig.~\ref{fig:The-growth-rate-1}
the growth rate given by Eq.~\ref{eq:grDA} is illustrated for $E/E_{c}=40$
and $60$ and $T_{e}=0.5$ keV. The growth rates from the LUKE calculations
are evaluated numerically, the Dreicer as a constant value ($\gamma_{D}$)
and the avalanche multiplication factor ($\bar{\gamma}_{A}$) from
the slope of the curve. The Dreicer growth rate calculated by LUKE
agrees well with predictions from Kulsrud (Ref.~\cite{kul73}) where
the Fokker-Planck equation is solved numerically. The avalanche multiplication
factor $\bar{\gamma}_{A}$ characterizes the tendency of a runaway
avalanche to develop, for a given magnetic equilibrium and parallel
electric field. The actual runaway production due to avalanche is
however time dependent since it is a product of the avalanche multiplication
factor $\bar{\gamma}_{A}$ and the time dependent runaway electron
density $n_{r}$. For example, $\bar{\gamma}_{A}$ can be non-zero,
even though the number of runaway electrons born from knock-on collisions
is negligible until a seed of primary electrons is established.

\begin{figure}
\centering
\includegraphics[scale=0.45]{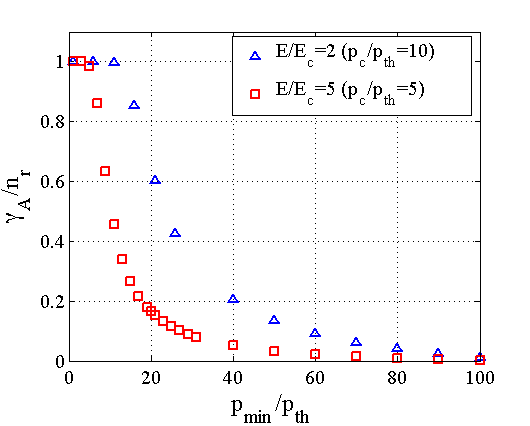}
\caption{Avalanche multiplication factor as a function of the lower momentum
cut off $p_{min}/p_{th}$ for $T_{e}=5$ keV, normalized to the avalanche
factor at $p_{min}=p_{c}$.\label{fig:Avalanche-multiplication-factor}}
\end{figure}

For the sake of simplicity, the Rosenbluth model in Ref.~\cite{ros97}
considers only secondary electrons born with a momentum larger than
$p_{c}$. However, electrons accelerated via a knock-on collision
to an intermediate momentum $p_{th}<p<p_{c}$ could contribute to
the runaway growth rate indirectly by populating the suprathermal
region and thereby modifying the Dreicer flux. Numerically, three
thresholds must be defined when implementing the Rosenbluth model
(\ref{eq:operator}): the minimum and maximum values for the secondary
electron momentum, and the minimum value $p_{re}$ above which runaways
are counted as primary electrons in the knock-on process. In order
to determine these parameters, the lower threshold above which knock-on
collisions are included is varied and the results are shown in Fig.
\ref{fig:Avalanche-multiplication-factor} for electric field $E/E_{c}=2$
and $E/E_{c}=5$ ($T_{e}=5$ keV). We can see that the indirect contribution
of knock-on collisions to suprathermal energies $p_{th}<p<p_{c}$
is negligible, such that it is appropriate to set the lower threshold
for secondary electron momentum at $p_{c}$. Energy conservation imposes that the higher threshold for secondary electron momentum is lower than $p_{re}$. 
We see that setting $p_{re}=4p_{c}$ is sufficient
to account for more than 80\% of knock-on collisions while ensuring
energy conservation.

\subsection{Bounce-averaged knock-on source term}

Since knock-on accelerated electrons emerge with high perpendicular
momentum \cite{ros97}, it is necessary to properly account for the
guiding-center dynamics in non-uniform magnetic field geometry and
treat the full 2-D momentum electron dynamics. In a non-uniform magnetic
field, the magnetic moment is an adiabatic invariant such that the
guiding center parallel velocity varies along the electron trajectory.
The pitch angle coordinate $\xi$ in Eq.~\ref{eq:operator} can be
expressed as a function of $(\xi_{0},\psi,\theta)$ where $\xi_{0}$
is the pitch angle measured at the poloidal position $\theta_{0}$
of the minimum magnetic field $B_{0}(\psi)$ on a magnetic flux surface.
When the collisional time is longer than the bounce period \cite{dec04a},
the rapid poloidal motion ensures that the electron distribution $f(\psi,p,\xi_{0})$
is independent of the poloidal angle $\theta$. The poloidal angle
dependence can thus be averaged out of the kinetic equation by bounce-averaging,
defined as

\begin{equation}
\{\mathbf{S}\}(\psi,p,\xi_{0})=\frac{1}{\lambda\tilde{q}}\left[\frac{1}{2}\sum_{\sigma}\right]_{T}\int_{_{_{\theta_{min}}}}^{^{^{\theta_{max}}}}\frac{d\theta}{2\pi}\frac{1}{|\hat{\psi}\cdot\hat{r}|}\frac{r}{R_{p}}\frac{B}{B_{p}}\frac{\xi_{0}}{\xi}\mathbf{S}(\psi,p,\xi),
\end{equation}
where $R_{p}$ is the major radius, $\theta_{min}$ and $\theta_{max}$
are the poloidal turning points for the trapped electrons, $B_{p}$
is the poloidal component of the magnetic field $B$ and the sum over
$\sigma$ applies to trapped particles $(T)$ only. The normalized
bounce time is 
\[
\lambda(\psi)=\frac{1}{\tilde{q}(\psi)}\int_{_{\theta_{min}}}^{^{\theta_{max}}}\frac{d\theta}{2\pi}\frac{1}{|\hat{\psi}\cdot\hat{r}|}\frac{r}{R_{p}}\frac{\xi_{0}}{\xi}\frac{B}{B_{p}},
\]
with
\[
\tilde{q}(\psi)\equiv\int_{_{0}}^{^{2\pi}}\frac{d\theta}{2\pi}\frac{1}{|\hat{\psi}\cdot\hat{r}|}\frac{r}{R_{p}}\frac{B}{B_{p}}
\]
 In the code LUKE, the electron distribution is normalized to a reference
density $n_{e}^{\dagger}$ and the time evolution is normalized to
the reference thermal collision frequency $\nu_{coll}^{\dagger}=1/\tau(v_{th})$,
so that the resulting source term is $\bar{\mathbf{S}}=\mathbf{S/S^{\dagger}}$
where $\mathbf{S}$ is from Eq.~\ref{eq:operator} and $\mathbf{S}^{\dagger}=n_{e}^{\dagger}\nu_{coll}^{\dagger}$
Momentum is normalized to the thermal momentum $\bar{p}=p/p_{th}$.
The knock-on source term is decomposed as $\mathbf{\bar{S}}(p,\psi,\xi,\theta)=\bar{\mathbf{S}}^{*}\delta(\xi-\xi^{*}(\bar{p}))$
where 
\begin{equation}
\bar{\mathbf{S}}^{*}=\frac{1}{4\pi}\frac{\beta_{th}^{\dagger2}}{\ln\Lambda^{\dagger}}\bar{n}_{e}\bar{n}_{r}\frac{1}{\bar{p}\gamma(\gamma-1)^{2}},
\end{equation}
is independent of $\theta$, so that $\{\bar{\mathbf{S}}\}=\mathbf{\bar{S}}^{*}\{\delta\left(\xi-\xi^{*}(\bar{p})\right)\}$
where $\xi$ is the pitch angle cosine at the poloidal angle position
$\theta$

\begin{equation}
\xi(\xi_{0},\psi,\theta)=\sigma\sqrt{1-\Psi(\psi,\theta)(1-\xi_{0}^{2})}.\label{eq:xi-1}
\end{equation}
Here $\Psi(\psi,\theta)=B(\psi,\theta)/B_{0}(\psi)$ and $\sigma=sign(v_{||})=sign(\xi_{0})$
indicates the direction of the electrons along the magnetic field
line. Using the general relation for Dirac's delta function $\delta(g(x))=\sum_{k}\delta(x-x_{k})/|g'(x_{k})|$
where $x_{k}$ are the zeros of the function $g(x)$ and $g'(x)=dg/dx$
provided that $g(x)$ is a continuously differentiable function and
$g'(x)$ is non-zero:

\begin{equation}
\delta(\xi-\xi^{*})=\sum_{k}\frac{2\delta(\theta-\theta_{k}^{*})|\xi^{*}|}{|\Psi'(\psi,\theta_{k}^{*})|(1-\xi_{0}^{2})},\label{eq:delta}
\end{equation}
where $\theta_{k}^{*}$ is the poloidal angle at which the secondary
electron emerges. From Eq.~\ref{eq:xi-1} $\theta_{k}^{*}$ is given
by

\begin{equation}
\sigma\sqrt{1-\Psi(\psi,\theta_{k}^{*})(1-\xi_{0}^{2})}-\xi^{*}=0,
\end{equation}
or

\begin{equation}
\Psi(\psi,\theta_{k}^{*})=\frac{B_{\theta_{k}^{*}}}{B_{0}}=\frac{1-\xi^{*2}}{1-\xi_{0}^{2}}=\frac{2}{(1-\xi_{0}^{2})(\gamma+1)}.\label{eq:phi}
\end{equation}
Using Eq.~\ref{eq:delta}, the delta function can be expressed as

\[
\{\delta(\xi-\xi^{*})\}=\frac{1}{\lambda\widetilde{q}}\frac{1}{\pi}\sum_{k}\frac{1}{|\hat{\psi}\cdot\hat{r}|_{\theta_{k}^{*}}}\frac{r_{\theta_{k}^{*}}}{R_{p}}\frac{B_{\theta_{k}^{*}}}{B_{p,\theta_{k}^{*}}}\frac{\xi_{0}}{\xi_{\theta_{k}^{*}}}\frac{|\xi^{*}|}{|\Psi'(\psi,\theta_{k}^{*})|(1-\xi_{0}^{2})}.
\]
and since $B_{\theta_{k}^{*}}=(1-\xi^{*2})/(1-\xi_{0}^{2})B_{0}$
with $\xi_{\theta_{k}^{*}}=\xi^{*}$

\begin{equation}
\{\delta(\xi-\xi^{*})\}=\frac{1}{\lambda\widetilde{q}}\frac{1}{\pi}\sum_{k}\frac{1}{|\hat{\psi}\cdot\hat{r}|_{\theta_{k}^{*}}}\frac{r_{\theta_{k}^{*}}}{R_{p}}\frac{B_{0}}{B_{p,\theta_{k}^{*}}}|\xi_{0}|\frac{(1-\xi^{*2})}{|\Psi'(\psi,\theta_{k}^{*})|(1-\xi_{0}^{2})^{2}},
\end{equation}
and the normalized, bounce-averaged avalanche operator becomes
\begin{eqnarray}
\{\bar{\mathbf{S}}(p,\psi,\xi_{0})\} & = & \frac{1}{2\pi^{2}}\frac{1}{\ln\Lambda^{\dagger}R_{p}}\bar{n}_{e}\bar{n}_{r}\cdot\nonumber \\
 & \times & \frac{1}{\bar{p}^{3}\gamma(\gamma-1)}\frac{B_{0}}{\lambda\widetilde{q}}\frac{|\xi_{0}|}{(1-\xi_{0}^{2})^{2}}\sum_{k}\Bigl[\frac{1}{|\hat{\psi}\cdot\hat{r}|}\frac{r}{B_{p}}\frac{1}{|\Psi'|}\Bigr]_{\theta_{k}^{*}},\label{eq:sbounce}
\end{eqnarray}
using the relation $p^{2}=(\gamma^{2}-1)=(\gamma-1)(\gamma+1)$ and
$p=\bar{p}\beta_{th}$.

\section{Effect of toroidicity\label{sec:Effect-of-toroidicity}}

The reduction of the Dreicer runaway rate away from the magnetic axis
has been identified in previous work \cite{eri03} including with
the code LUKE, which solves the bounce-averaged kinetic equation \cite{dec04a}.
At least three effects contribute to reduce the growth rate : the
overall effect of the electric field on trapped electrons cancels
out over one bounce period; the acceleration of passing electrons
is also reduced as their pitch angle increases towards the high field
side; the existence of a magnetic trapping cone creates larger pitch-angle
gradients in the circulating region, thereby increasing the effect
of pitch-angle scattering.

As discussed in Sec.~\ref{sec:Knock-on-collisions}, secondary electrons
emerging from the knock-on collisions are typically highly magnetized.
Since the trapping effect increases off the magnetic axis in a non-uniform
magnetic field configuration, the further away from the magnetic axis
the electrons appear, the more they tend to be born trapped \cite{ros97}. 

\begin{figure}
\centering
\includegraphics[scale=0.4]{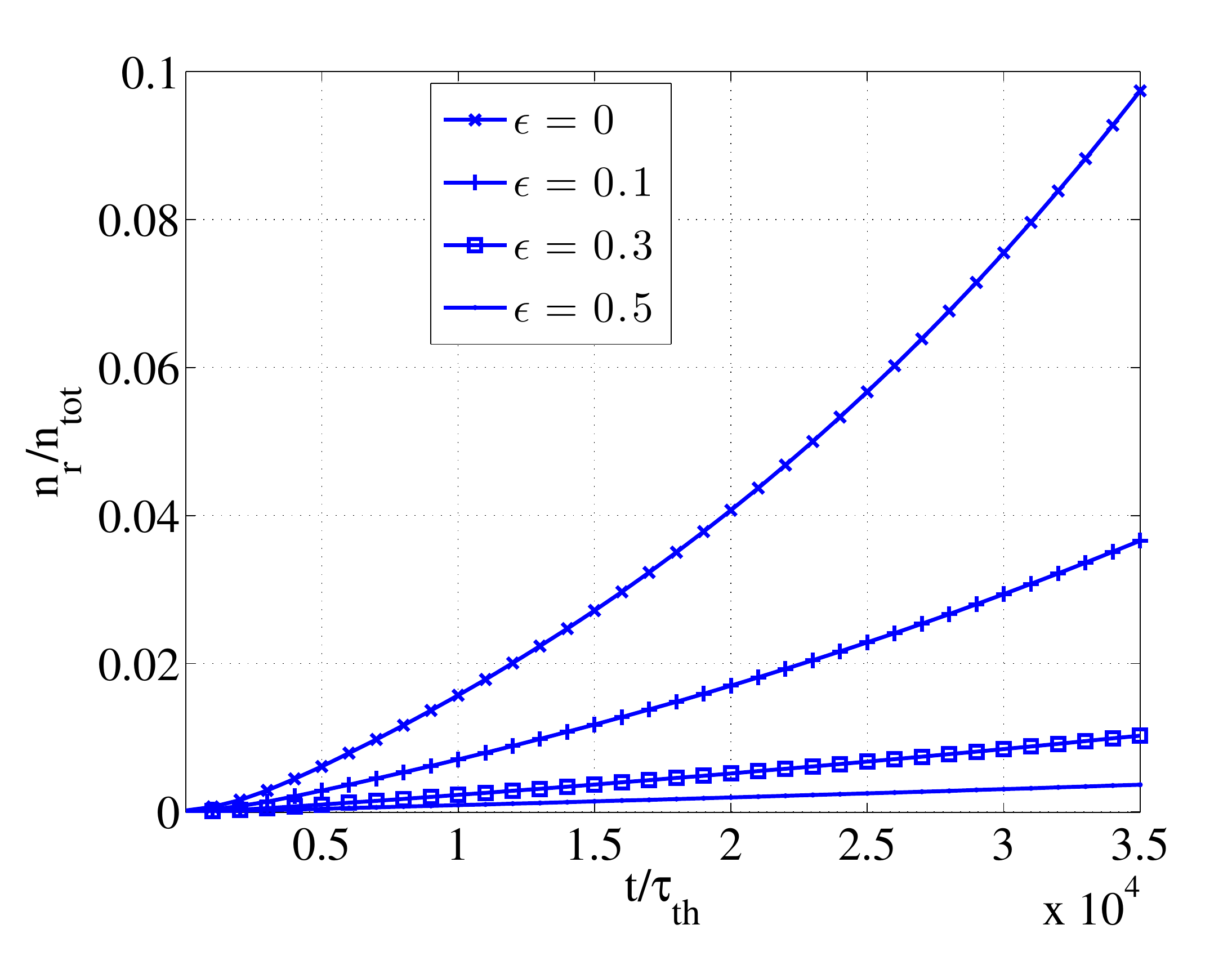}
\caption{The evolution of the runaway electron population, including the avalanche
effect owing to knock-on collisions, depends strongly on the radial
position in a non-uniform magnetic field configuration, where $\epsilon=r/R$
is the inverse aspect ratio coordinate.\label{fig:The-evolution-of}
$E/E_{c}=40$, $T_{e}=0.5$ keV and the time $t$ is normalized to
the thermal collision time $\tau_{th}$.}
\end{figure}

To quantify the tendency of magnetic trapping, the evolution of the
runaway population is calculated in a scenario with circular plasma
cross section and magnetic non-uniformity, with inverse aspect ratio
ranging from $\epsilon=0$ to $\epsilon=a/R=1$. The inverse aspect
ratio of the Tore Supra tokamak is $\epsilon\approx0.3$. The calculations
in Fig.~\ref{fig:The-evolution-of} reveal that the runaway electron
population grows significantly slower off the magnetic axis than in
the center. 

\begin{figure}
\centering
\includegraphics[scale=0.45]{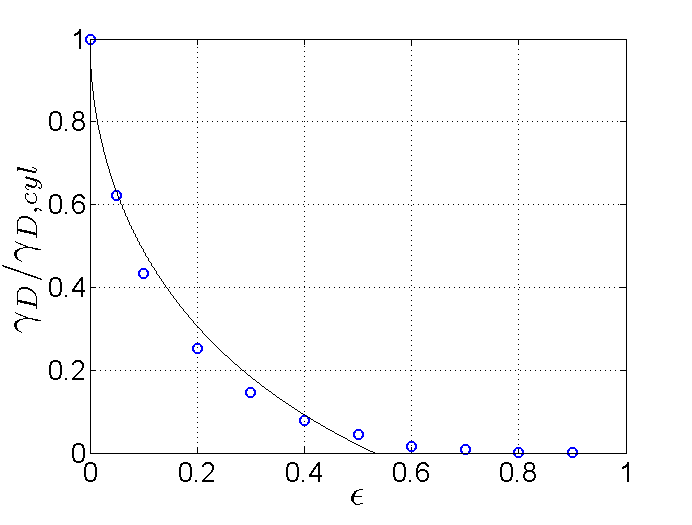}
\caption{Radial dependence of Dreicer growth rate, normalized to the growth
rate for cylindrical case $\epsilon=0$ and a fit $\gamma_{D}/\gamma_{D,cyl}=1-1.2\sqrt{2\epsilon/(1+\epsilon)}$.\label{fig:Radial-dependence-of}}
\end{figure}

In order to study the trapping effects on the runaway population,
the Dreicer growth rate $\gamma_{D}$ and the avalanche multiplication
factor $\bar{\gamma}_{A}$ are calculated with the bounce-averaged
code LUKE and quantified separately. The Dreicer growth rate is found
to be strongly affected by the non-uniformity of the magnetic field,
as shown in Fig.~\ref{fig:Radial-dependence-of}. A fit of the numerical
results gives an analytic expression of the Dreicer growth rate $\gamma_{D}/\gamma_{D,cyl}=1-\sqrt{2\epsilon/(1+\epsilon)}$.
The results indicate that for $\epsilon>0.5$ runaway generation from
Dreicer acceleration vanishes.

\begin{figure}
\centering
\includegraphics[scale=0.45]{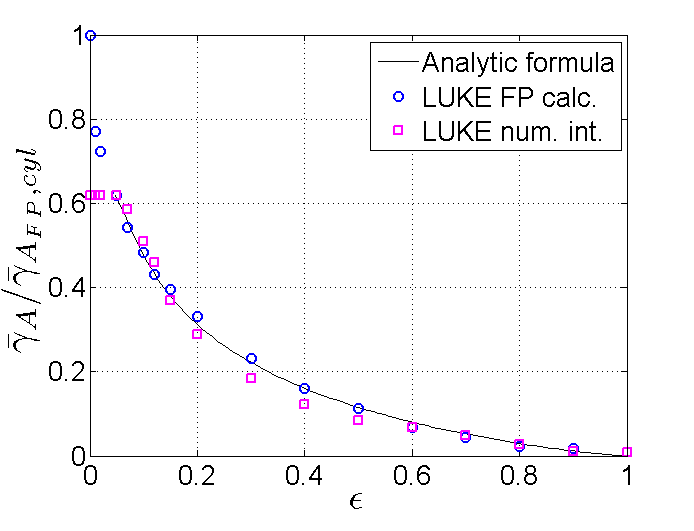}
\caption{Radial dependence of the avalanche multiplication factor from bounce-averaged
LUKE calculations (circles), normalized to to the avalanche multiplication
factor for the cylindrical case $\epsilon=0$.\label{fig:Radial-dependence-of-1}
The numerical integration over the knock-on source term in Eq.~\ref{eq:operator}
with the toroidal dependence in the momentum integration boundary
is plotted in squares. The solid line shows the analytic estimate
of the growth rate off the magnetic axis from Eq.~\ref{eq:grcirc-1}.}
\end{figure}

A reduction of $\bar{\gamma}_{A}$ away from the magnetic axis is
observed in Fig.~\ref{fig:Radial-dependence-of-1}, with an avalanche
multiplication factor that decreases with the inverse aspect ratio.
In order to derive an analytical estimate for the avalanche growth
rate including the effect of magnetic trapping owing to a non uniform
magnetic configuration, it is assumed that all electrons with momentum
$p>p_{c}$ will contribute to the runaway population (as in Ref.~\cite{ros97}),
except the secondary electrons that appear in the trapped momentum
region $p<p_{T}$. The magnetic trapping criterion on the momentum
$p_{T}$ of secondary electrons born via knock on collisions is 

\begin{equation}
\frac{B(\theta)}{B_{max}}>\frac{2}{\sqrt{1+p_{T}^{2}}+1},\label{eq:trap}
\end{equation}
where $B_{max}/B(\theta)=(1+\epsilon\cos\theta)/(1-\epsilon)$ in
a circular concentric magnetic configuration. Electrons will run away
if their momentum exceeds both the critical momentum and the trapping
condition in Eq.~\ref{eq:trap}. The lower integration limit $p_{min}$
for the analytical estimate of the avalanche growth rate is thus given
by max$(p_{c},p_{T})$. An analytical expression for the inverse aspect
ratio dependent avalanche growth rate is obtained by integrating the
source term from over momentum space from $p_{min}$ to $p_{max}=\infty$
, which results in a locally modified, inverse aspect ratio dependent
avalanche growth rate

\begin{eqnarray}
\frac{dn_{r}}{dt_{n}}(\theta,\epsilon) & = & \frac{1}{2}\frac{1}{\ln\Lambda^{\dagger}}\bar{n}_{e}\bar{n}_{r}\frac{1}{\sqrt{1+p_{min}^{2}}-1}\nonumber \\
 & = & \frac{1}{2}\frac{1}{\ln\Lambda^{\dagger}}\bar{n}_{e}\bar{n}_{r}\:\mathsf{min}\left(\frac{E}{E_{c}},\frac{\left(1-\epsilon\right)^{2}}{2\epsilon(1+cos\theta)}\right).\label{eq:dndtepsi}
\end{eqnarray}

The flux surface averaged growth rate is derived in Appendix \ref{sec:Appendix}.
For $\epsilon E/E_{c}\gg1$, $\theta_{b}\rightarrow\pi$ , the growth
rate is reduced by a factor $(1-\epsilon)^{2}/\left(\pi\sqrt{\epsilon E/E_{c}}\right)$.
The inverse aspect ratio dependence of the estimated avalanche growth
rate obtained from Eq.~\ref{eq:dndtepsi} is compared to numerical
results. In addition, a numerical integration of the source term is
also performed, with the same criterion on the lower integration boundary
in momentum space $p_{min}$ as the analytic estimate in Eq.~\ref{eq:grcirc-1}
. The analytic result is also compared to avalanche growth rate from
Fokker-Planck calculations with the LUKE code. In that case, the trapping
conditions are the same as in the analytic result, except for that
the critical momentum is pitch angle dependent $p_{c}^{2}=E_{c}/\left(E\xi\right)$.
The LUKE calculated avalanche multiplication factor and the analytical
estimate show good agreement (Fig.~\ref{fig:Radial-dependence-of-1}).

Figure \ref{fig:Radial-dependence-of-1} shows the reduced growth
rate for $E/E_{c}=5$, relative to a cylindric plasma, equivalent
to the growth rate on the magnetic axis ($\epsilon=0$). Numerical
integration of the source term shows good agreement with the analytic
result (Eq.~\ref{eq:grcirc-1}). Close to the center, at low inverse
aspect ratio, the effect of trapping is not visible, since the critical
momentum is higher than the trapped momentum over the whole flux surface.
This effect decreases with increasing $E/E_{c}$ as the critical momentum
$p_{c}$ decreases and becomes less restrictive compared to the trapping
condition $p_{T}$, which explains the flat top seen in Fig.~\ref{fig:Radial-dependence-of-1}.
However, for the FP calculations the magnetic trapping effect influences
the growth rate also close to the magnetic axis. A possible explanation
is pitch angle collisions that couple the dynamics between the trapped
and the passing region.

The growth rate obtained from bounce-averaged calculations suggest
that the formation of runaway electrons is slower the further away
from the magnetic axis they appear. In other words, the time scale
of the local growth rate could be longer than suggested by collisional
theory \cite{kul73,ros97}. Potential loss mechanisms, such as transport
of fast electrons due to magnetic field perturbations \cite{zen13}
could therefore act more efficiently on the runaway electrons formed
off the magnetic axis than the ones formed on axis which could lead
to well confined runaway electrons at the center of the plasma. 

\section{The relative importance of the avalanche effect\label{sec:Importance-of-avalanche}}

The results presented in Sec.~\ref{sec:Runaway-electron-growth} (see
Fig.~\ref{fig:The-runaway-population}) have shown that the runaway
electron distribution can be significantly modified by including the
effect of knock-on collisions. In order to understand the mechanisms
that govern the runaway electron generation processes a parametric
study is performed with the aim to investigate which runaway formation
process, Dreicer or avalanche, dominates in non-disruptive tokamak
experiments.

The relative importance of the avalanche mechanism to the Dreicer
mechanism can be estimated by comparing the analytic avalanche growth
rate in Eq.~\ref{eq:grRP-1} and the Dreicer generation that is derived
in Ref.~\cite{con75}:

\[
\left(\frac{\partial n_{r}}{\partial t}\right)_{D}\sim\frac{2}{\sqrt{\pi}}n_{e}\nu(v_{th})\left(\frac{E}{E_{D}}\right)^{1/2}\exp\left(-\frac{E_{D}}{4E}-\left(\frac{2E_{D}}{E}\right)^{1/2}\right),
\]
where $E_{D}=\left(c/v_{th}\right)^{2}E_{c}$ is the electric field
at which even thermal electrons will run away, known as the Dreicer
field. The ratio of the two growth rates is

\begin{equation}
\frac{\gamma_{A}}{\gamma_{D}}\sim\frac{\sqrt{\pi}}{4}\frac{n_{r}}{n_{e}}\frac{1}{\ln\Lambda}\left(\frac{v_{th}}{c}\right)^{3}\left(\frac{E}{E_{c}}-1\right)\left(\frac{E}{E_{D}}\right)^{-1/2}\exp\left(\frac{E_{D}}{4E}+\sqrt{\frac{2E_{D}}{E}}\right).\label{eq:nand}
\end{equation}

\begin{figure}
\centering
\includegraphics[scale=0.38]{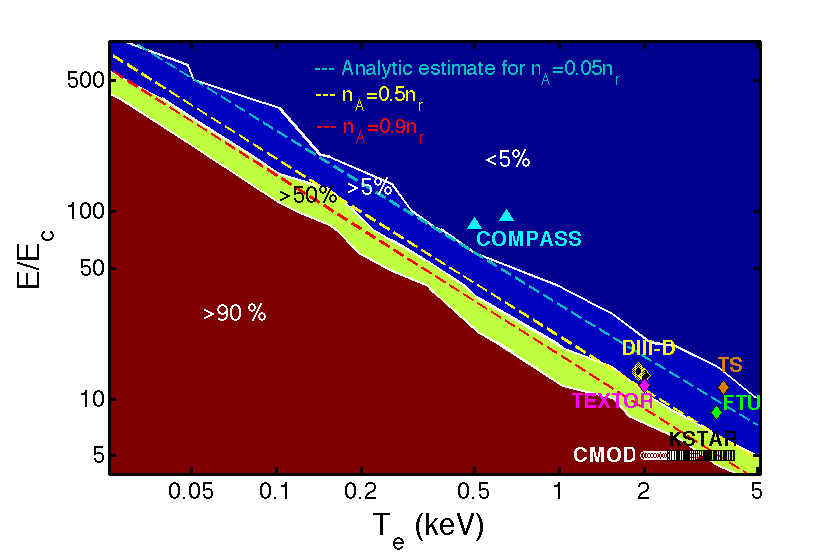}
\caption{The fraction of runaways originating from knock-on collisions $(n_{A}/n_{r})$
as modelled in LUKE. The analytic estimate of when $5\%$ (cyan line),
$50\%$ (yellow line) and $90\%$ (red line) of the runaways come
from avalanche is obtained from Eq.~\ref{eq:nand}. Relation to non-disruptive
scenarios where runaway electrons were generated in several tokamaks.
All data points are taken from Ref.~\cite{gra14} except for the Tore
Supra (TS) point (discharge $\#40719$) and COMPASS points (discharge
$\#8555$ and $\#8630$).\label{fig:The-significance-of}}
\end{figure}

\begin{figure}
\centering
\includegraphics[scale=0.38]{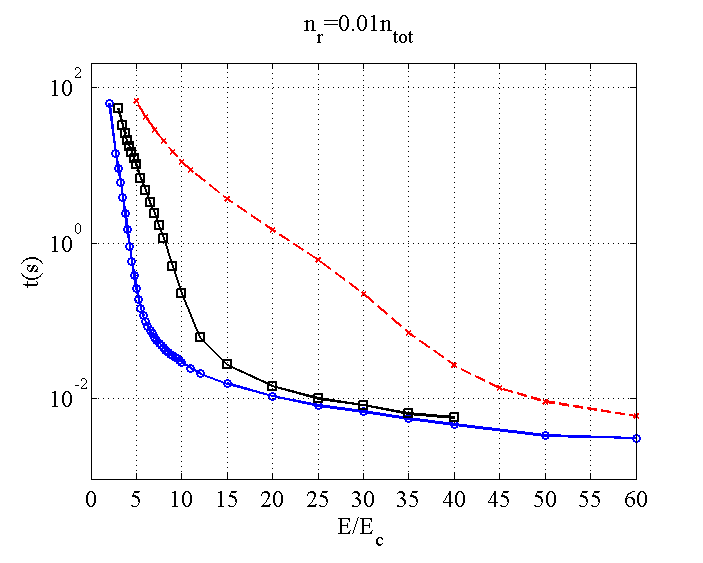}\protect\caption{The time required for $1\%$ of the Maxwellian electrons to run away,
for the electron temperature $T_{e}=\,0.5$ keV (dashed line), $2$
keV (solid line and squares) and $5\,\textrm{keV}$ (solid line and
circles).\label{fig:The-time-required}}
\end{figure}

By letting a small fraction of electrons run away in LUKE, the relative
importance of the avalanche effect as a function of plasma temperature
and electric field strength can be evaluated numerically from the
fraction of the runaway electrons that originate from Dreicer and
knock-on collisions. In Fig.~\ref{fig:The-significance-of} the fraction
of runaway electrons born from knock-on collisions is shown, when
$1\%$ of the initial electron population has run away in a cylindrical
magnetic configuration with constant electric field, density and temperature.
The fraction of runaway electrons has to be small enough for the equilibrium
parameters to remain constant. The relative importance of secondary
runaway electrons grows at lower temperature and electric field, as
the slower primary generation in high collisionality (low temperature)
allows for runaway avalanches to take off. The time required to reach
the runaway fraction varies strongly in the parameter space presented
in Fig.~\ref{fig:The-significance-of}. The time required for $1\%$
of the electrons to run away is illustrated for various electron temperatures
($T_{e}=$ $0.05,\,0.5,\,2$ and $5\,\textrm{keV}$) in Fig.~\ref{fig:The-time-required}.
The formation of runaway electrons slows down as the collisionality
increases at lower bulk temperature.

The numerical results are compared with the analytical estimate from
Eq.~\ref{eq:nand} with $n_{r}/n_{e}=0.01$. The condition for the
dominance of the avalanche effect $\gamma_{A}/\gamma_{D}>1$ is plotted
in Fig.~\ref{fig:The-significance-of} along with the boundaries for
which $n_{A}/n_{r}=5\%$ and $90\%$. 

In order to relate the study to experimental tokamak scenarios, it
must be noted that the simulations are performed for constant electric
field and temperature. Consequently, the study is restricted to non-disruptive
scenarios with well-diagnosed and quiescent conditions from several
tokamaks, where runaway electrons have been observed in the current
flattop with the relevant plasma parameters maintained essentially
constant. Results from scenarios with reproducible measurements of
electron density, loop voltage and plasma temperature at the runaway
electron onset from DIII-D, FTU, TEXTOR, Alcator C-Mod and KSTAR were
recently published in Ref.~\cite{gra14}. From this study the threshold
electric field normalized to the critical field is found to be significantly
higher than predicted by collisional theory where the birth of runaway
electrons is predicted at $E/E_{c}>1$, provided that no additional
runaway electron loss mechanisms are present \cite{kul73}. However,
the condition for runaway onset in collisional theory does not take
the time required to generate runaway electrons into account. Estimations
from LUKE calculations in Fig.~\ref{fig:The-time-required} shows
that this time scale can be unrealistically large as compared to the
tokamak discharge duration. The time to generate a small fraction
of runaway electrons from a Maxwellian distribution is finite for
$E/E_{c}>1$ but as $E/E_{c}\rightarrow1$, the required time to generate
runaway electrons $t\rightarrow\infty$. However, it is not the only
explanation since the study in Ref.~\cite{gra14} found that the $E/E_{c}$
threshold for suppression is also well above unity.

\begin{figure}[t]
\subfigure{ \includegraphics[scale=0.45]{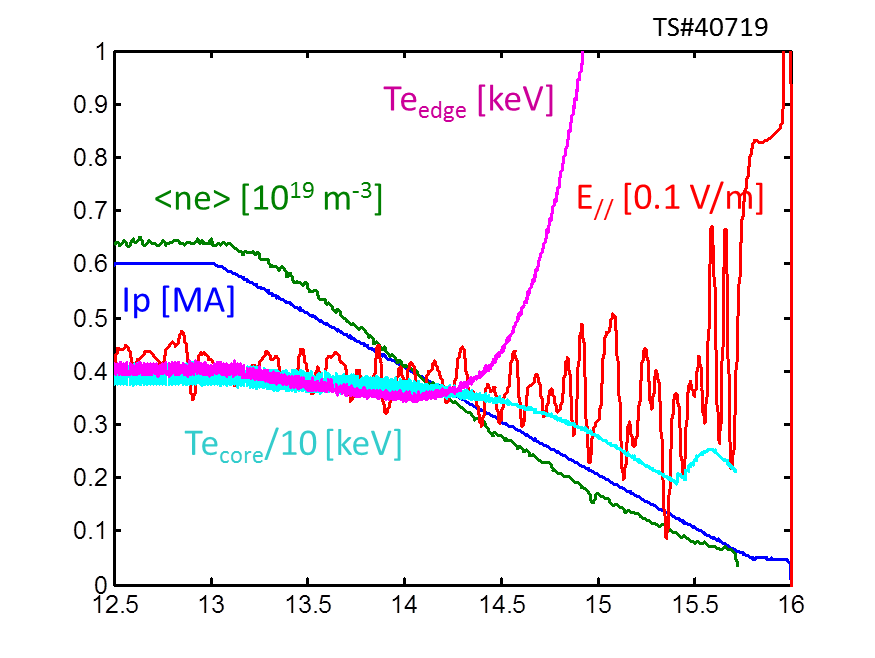} \label{fig:te19} }
\subfigure{ \includegraphics[scale=0.45]{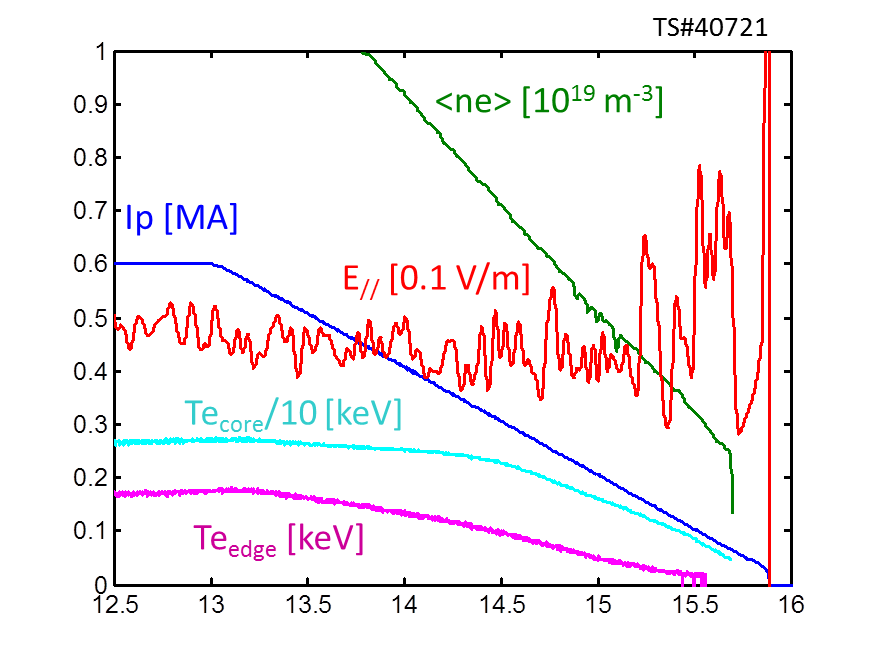}\label{fig:te21} }
\caption{Signature of suprathermal electrons
on the edge ECE chord at around $t=14.5$ s are seen in the Tore Supra
discharge $47019$ (left). In a following discharge, with higher density
(right), there is no sign of suprathermal electrons. \label{fig:Signature-of-suprathermal} }
\end{figure}

\begin{figure}
\centering
\includegraphics[scale=0.5]{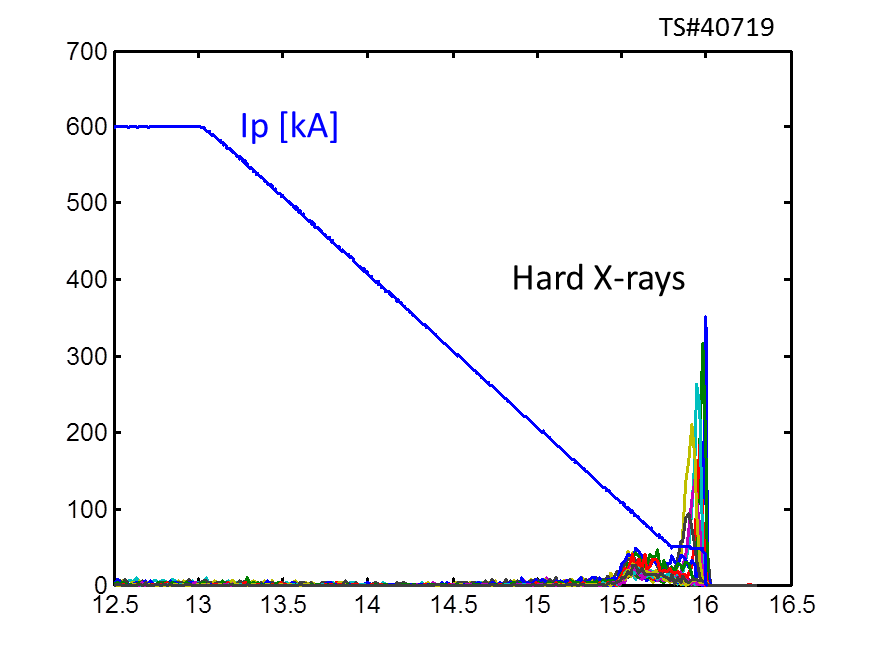}
\caption{HXR data from the vertical camera (channels 1-21) in the energy range
$E_{HXR}=20-200$ keV. The HXR emission produced in the current ramp
down in $40719$ is a signature of suprathermal electrons, whereas
in the higher density discharge $40721$ no HXR emission is detected.\label{fig:HXR}}
\end{figure}

Runaway electrons have been generated in the Tore Supra (TS) tokamak
in low density discharges ($n_{e}<10^{19}$ $m^{-3}$). The TS pulses
$\#40719$ and $\#40721$ are performed after a boronization and suprathermal
electrons are observed in the former discharge but not in the latter.
Both are ohmic discharges at $I_{p}=0.6$ MA in the current flattop.
Suprathermal electrons are observed in $\#40719$ by the ECE edge
chords at current ramp-up and ramp-down, when the density is low ($<n_{e}>=0.4\cdot10^{19}$
$m^{-3}$), see Fig.~\ref{fig:te19}. The uniform E-field, estimated
as the time derivative of the resistive flux \cite{eji82}, is $E_{\|}=0.038\pm0.003$
V/m and the core temperature is $3.8$ keV. The determination of the
magnetic flux at the plasma boundary is described in Ref.~\cite{wij97}.
No suprathermal electrons are detected by electron cyclotron emission
(ECE) in the following pulse \#$40721$ at a higher electron density,
see Fig.~\ref{fig:te21}. Similar result is found from HXR measurements
from the vertical camera detecting emission of $20-200$ keV (Fig.
\ref{fig:HXR}). A peak of photo-neutrons is observed at the plasma
termination for the lower density shot ($\#40719$) but not for the
higher density shot ($\#40721$). From the combined observations on
ECE, HXR and photo-neutron measurements, the presence of relativistic
electrons during the ramp-down of $\#40719$ is identified. During
the current flattop of $\#40719$, the electron density is $<n_{e}>=0.64\cdot10^{19}$
$m^{-3}$, corresponding to $E/E_{c}\approx8$, but there is no sign
of suprathermals until $E/E_{c}\approx11$. The suprathermal generation
in $\#40719$ is added to the ($E/E_{c},T_{e}$) scan (see Fig.~\ref{fig:The-significance-of})
and lands in the region where Dreicer generation is dominant. In the
higher density pulse (\#$40721$) $E/E_{c}\approx4$ during the current
flattop and no suprathermal electrons are detected. These results
are in line with those of Ref.~\cite{gra14} where $E/E_{c}\sim3-12$
is required to generate a detectable population of runaway electrons
in the various tokamaks. 

Relating the data from the experiments in Ref.~\cite{gra14} and the
TS discharge $\#40719$ to the parameter scan done in LUKE (Fig.~\ref{fig:The-significance-of})
reveals that the scenarios fall in or close to the region where the
avalanche mechanism becomes significant for the runaway electron growth
rate (Fig.~\ref{fig:The-significance-of}). Data from two COMPASS
discharges where runaways were observed ($\#8555$ and $\#8630$)
fall in the region where the Dreicer effect is dominant \cite{vla15}.
Runaway electrons are commonly produced in the current ramp-up phase
of the COMPASS tokamak, due to the relatively high $E/E_{c}$ ratio
($20-200$). The circular $130$ kA discharge \#$8555$ was part of
the electron density $<n_{e}>$ scan from $1-4\cdot10^{19}$ $m^{-3}$,
where $<n_{e}>$ for this particular shot was $2\cdot10^{19}$ $m^{-3}$
during the flattop. The rise in runaway activity was observed with
HXR NaI(Tl) scintillator and photoneutron detector as the $<n_{e}>$
decreased from discharge to discharge, while Parail-Pogutse instability
appeared for all discharges with $<n_{e}>$ lower than in the discharge
\#8555. D-shaped $160$ kA discharge \#8630 was done for the purpose
of the sawteeth-runaway correlation studies with the electron density
$<n_{e}>=9\cdot10^{19}$ $m^{-3}$. Even though the discharge had
relatively high $<n_{e}>$, the runaway activity correlated with the
sawteeth instability was visible in HXR and photoneutron signals.
These two COMPASS discharges \#8555 and \#8630 are plotted on Fig.
\ref{fig:The-significance-of}, where $E/E_{c}$ at the ramp-up phase
were $85$ and $94$, respectively. The electron density at the time
of the runaway detection is $n_{e}=1.1$ and $0.80$ $m^{-3}$. In
COMPASS, interferometry is used for the line averaged electron density
$<n_{e}>$ measurements, while Thomson scattering is used for electron
temperature $T_{e}$ and electron density $n_{e}$ profile measurements. 
These observations suggest that knock-on collisions may contribute
to the formation of runaway electron generation in tokamak plasmas,
even in non-disruptive scenarios. The study predicts that avalanches
can play an important role during current flattop. A self consistent
electric field and equilibrium solver would be necessary to study
avalanches with LUKE in disruptions, but is beyond the scope of the
current work.

\section{Conclusion}

In this work the growth of runaway electron populations through the
combined effect of Dreicer and knock-on collision mechanisms is studied.
The Rosenbluth \cite{ros97} model is extended to non-uniform magnetic
field configurations and implemented as a bounce-averaged conservative
source/sink term within the kinetic equation in the 3-D Fokker-Planck
solver LUKE. Dependencies of key parameters such as electric field
strength, electron temperature, and density are investigated. In addition,
magnetic trapping effects are quantified in a non-uniform magnetic
equilibrium, resulting in a reduced runaway population off the magnetic
axis for both the Dreicer and the avalanche mechanism.

The kinetic modelling of the formation of runaway electrons is restricted
to non-disruptive scenarios as found in the current flattop with non-transient
electric field and plasma temperature. Modelling the rapidly varying
temperature and electric field found in disruptions would require
a proper description of the thermal quench with implemented radiative
or convective loss mechanisms of the plasma energy, including MHD
instabilities. The extension of the kinetic code LUKE to 3-D magnetic
configuration and its coupling with a fluid code such as JOREK \cite{huy07}
would be necessary for such a purpose and is beyond the scope of this
work. In the present paper, runaways are confined to the flux-surface
where they are generated, such that the growth rate derived herein
should be considered as upper estimates.

Since knock-on accelerated electrons emerge with high perpendicular
momentum, the full 2-D guiding-center momentum dynamics is taken into
account via a bounce-averaged description. The effect of magnetic
trapping of the electrons in a non-uniform magnetic field configuration,
known as the magnetic mirror effect, has been investigated, revealing
reduction of both Dreicer and avalanche mechanisms off the magnetic
axis. An analytical expression for avalanche growth rate accounting
for magnetic trapping is derived. It is in agreement with numerical
simulations and shows that a significant proportion of secondary electrons
are knocked into the trapped region off the magnetic axis. The reduction
of the off axis growth rate implies that the time scale of runaway
formation is longer at the edge than close to the center, which means
that potential loss mechanisms such as radial electron transport could
compete with the acceleration of runaway electrons at the edge.

Moreover, quantifying the relative importance of avalanche generation
as a function of plasma temperature and electric field strength, the
simulations reveal that runaway electrons originating from knock-on
collisions dominate at low temperature and electric field strength
and likely play a significant role in runaway generation processes
in several tokamaks with data from non-disruptive scenarios that are
presented in Ref.~\cite{gra14}. The onset of runaway electrons found
in these experiments is related to LUKE simulations of corresponding
electric field and temperature in order to evaluate the importance
of the avalanche effect, revealing that knock-on collisions may play
a significant role also in non-disruptive scenarios. The LUKE calculations
predict runaway electron generation also in a near critical field,
in agreement with collisional theory if no other runaway electron
loss mechanisms than collisional damping are present. However, the
time to generate runaway electrons can be significantly large compared
to the duration of the phase in which $E/E_{c}>1$ in experiments.
In addition, the required time for runaway electron formation is very
sensitive to the bulk electron temperature. The lack of runaway electron
signatures near the critical electric field could therefore be explained
by the long time scale required for their formation. To understand
this discrepancy between observations and theory, the existence of
additional loss mechanisms that dominate during the current flattop
must be addressed. One possible candidate is transport of fast electrons
due to magnetic field perturbations \cite{zen13}. Once such additional
runaway electron loss mechanisms have been identified, the LUKE code
may form an excellent test bed for quantifying these effects, which
will be the objective of future work.

\paragraph*{Acknowledgement}

This work has been carried out within the framework of the EUROfusion
Consortium and has received funding from the Euratom research and
training programme 2014-2018 under grant agreement No 633053. The
views and opinions expressed herein do not necessarily reflect those
of the European Commission.

\section*{References}
\bibliographystyle{unsrt}

\clearpage

\appendix
\section{ \label{sec:Appendix}}

\section*{Derivation of toroidicity dependent avalanche growth rate }

As described in Sec.~\ref{sec:Effect-of-toroidicity}, the avalanche
growth rate is evaluated by the flux surface averaged knock-on source
term in Eq.~\ref{eq:operator} where the lower integration boundary
is set by the maximum of the critical momentum $p_{c}$ and the momentum
defining the boundary of a passing and a trapped electron $p_{T}$,
given by the trapping condition in Eq.~\ref{eq:trap}. For finite
$E/E_{c}$, the critical momentum $p_{c}>0$ . As the growth rate
is averaged over the poloidal angle, $p_{min}\rightarrow p_{c}$ as
the high field side is approached ($p_{T}\rightarrow0$ as $\theta\rightarrow\pi$).
The growth rate becomes: 
\begin{eqnarray}
\frac{dn_{r}}{dt_{n}}(\theta,\epsilon) & = & \frac{1}{2}\frac{1}{\ln\Lambda^{\dagger}}\bar{n}_{e}\bar{n}_{r}\frac{1}{\sqrt{1+p_{min}^{2}}-1}=\label{eq:dndtepsi-1}\\
 &  & =\frac{1}{2}\frac{1}{\ln\Lambda^{\dagger}}\bar{n}_{e}\bar{n}_{r}\:\mathsf{min}\left(\frac{E}{E_{c}},\frac{\left(1-\epsilon\right)^{2}}{2\epsilon(1+cos\theta)}\right).
\end{eqnarray}
The poloidal angle $\theta_{bound}$ where $p_{c}=p_{T}$ constitutes
the boundary between the region where the avalanche rate is limited
either by the drag force or by the magnetic trapping effect. This
angle is obtained from the condition $p_{c}=p_{T}$:
\begin{eqnarray*}
1+cos\theta_{bound} & = & (1-\epsilon)^{2}/(2\epsilon\frac{E}{E_{c}})\rightarrow\\
 &  & \theta_{bound}=\pm\arccos((1-\epsilon)^{2}/(2\epsilon\frac{E}{E_{c}})-1).
\end{eqnarray*}
If $\epsilon E/(E_{c}(1-\epsilon)^{2})<1/4$, $p_{c}$ is the lower
integration limit $p_{min}$ and if $\epsilon E/(E_{c}(1-\epsilon)^{2})>1/4$,
$p_{min}=p_{T}(\theta)$. Averaged over the flux surface according
to volumic flux surface average the growth rate is:

\begin{eqnarray}
 & < & \frac{dn_{r}}{dt_{n}}>_{V}(\epsilon)=\label{eq:grFS-1}\\
 &  & =\frac{1}{\hat{q}}\left(\frac{1}{\pi}\int_{0}^{\theta_{bound}}\epsilon\frac{B_{0}(\epsilon)}{B_{P}}\frac{dn_{r}}{dt_{n}}(p_{T}(\theta))d\theta+\frac{1}{\pi}\int_{\theta_{bound}}^{\pi}\epsilon\frac{B_{0}(\epsilon)}{B_{P}}\frac{dn_{r}}{dt_{n}}(p_{c})d\theta\right)=\nonumber \\
 &  & =\frac{1}{2}\frac{1}{\ln\Lambda^{\dagger}}\bar{n}_{e}\bar{n}_{r}\times\nonumber \\
 &  & \times\left((1+\epsilon)\frac{B_{p}}{B}\frac{1}{\pi}\int_{0}^{\theta_{bound}}\frac{B_{0}(\epsilon)}{B_{P}}\frac{\left(1-\epsilon\right)^{2}}{2\epsilon(1+cos\theta)}d\theta+(1+\epsilon)\frac{E}{E_{c}}\frac{1}{\pi}\int_{\theta_{bound}}^{\pi}\frac{(1+\epsilon\cos(\theta))}{1+\epsilon}d\theta\right).\nonumber 
\end{eqnarray}
In the above calculation circular concentric flux surfaces are considered
so that $|\hat{\psi\cdot}\hat{r}|$=1, $r/R_{p}=\epsilon$ and 
\[
\hat{q}=\intop_{0}^{2\pi}\frac{d\theta}{2\pi}\epsilon\frac{B_{0}}{B_{p}}=\intop_{0}^{2\pi}\frac{d\theta}{2\pi}\epsilon\frac{(1+\epsilon\cos(\theta))}{(1+\epsilon)}\frac{B}{B_{p}}=\frac{\epsilon}{(1+\epsilon)}\frac{B}{B_{p}}.
\]
The flux surface averaged growth rate takes the form: 
\begin{eqnarray}
 & < & \frac{dn_{r}}{dt_{n}}(\theta,\epsilon)>_{V}=\frac{1}{2\ln\Lambda^{\dagger}}\bar{n}_{e}\bar{n}_{r}\frac{E}{E_{c}}\times\left(1-\frac{\theta_{bound}}{\pi}-\frac{\epsilon}{\pi}\sin(\theta_{bound})\right)\nonumber \\
 &  & +(1-\epsilon)^{2}\frac{1}{2\epsilon\pi}\left((1-\epsilon)\tan\left(\theta_{bound}/2\right)+\epsilon\theta_{bound}\right)\label{eq:grcirc-1}\\
 & = & \frac{1}{2\ln\Lambda^{\dagger}}\bar{n}_{e}\bar{n}_{r}\frac{E}{E_{c}}\times\nonumber \\
 &  & \times\left(1-\frac{\theta_{bound}}{\pi}-\frac{\epsilon}{\pi}\sin(\theta_{bound})+\frac{(1-\epsilon)^{2}E_{c}}{2\epsilon\pi E}\left(\sqrt{1-\epsilon}\sqrt{4\epsilon E/E_{c}-(1-\epsilon)}+\epsilon\theta_{bound}\right)\right),\nonumber 
\end{eqnarray}
 where 
\[
\tan\left(\theta_{bound}/2\right)=\frac{\sin(\theta_{bound})}{1+cos(\theta_{bound})}=\frac{\sqrt{4\epsilon E/E_{c}-(1-\epsilon)}}{\sqrt{1-\epsilon}}.
\]
For $\epsilon E/E_{c}\gg1$, $\theta_{b}\rightarrow\pi$ and the growth
rate is reduced by a factor $(1-\epsilon)^{2}/\left(\pi\sqrt{\epsilon E/E_{c}}\right)$.

\end{document}